\documentstyle{article}
\begin{document}

\noindent Stockholm\\
USITP 03-12\\
December 2003\\

\vspace{1cm}

\begin{center}

{\Large THE IMPORTANCE OF BEING UNISTOCHASTIC}\footnote{Talk at the 
Second V\"axj\"o conference on Quantum Theory: Reconsideration of 
Foundations, June 2003.} 

\vspace{1cm}

{\large Ingemar Bengtsson}\footnote{Email address: ingemar@physto.se. 
Supported by VR.}

\vspace{8mm}

{\sl Stockholm University, AlbaNova\\
Fysikum\\
S-106 91 Stockholm, Sweden}

\vspace{8mm}

{\bf Abstract}

\end{center}

\vspace{5mm}

\noindent A bistochastic matrix is a square matrix with positive entries such that 
rows and columns sum to unity. A unistochastic matrix is a bistochastic 
matrix whose matrix elements are the absolute values squared of a unitary 
matrix. We can now ask questions such as when a given bistochastic matrix is 
unistochastic. I review these questions: Why they are asked, why 
they are difficult to answer, and what is known about them.





  



\section{The problem}
There are some people that you have never heard of, but once you have 
met them for the first time they turn up everywhere. Unistochastic matrices 
are like that. First, some definitions: An $N \times N$ matrix $B$ is 
said to be {\it bistochastic} if its matrix elements obey

\begin{equation} \mbox{i}: \ B_{ij} \geq 0 \hspace{6mm} \mbox{ii}: 
\ \sum_iB_{ij} = 1 \hspace{6mm} 
\mbox{iii}: \ \sum_jB_{ij} = 1 \ . \label{1} \end{equation}

\noindent The first condition ensures that positive vectors are 
transformed to positive vectors. The second condition ensures that 
the sum of the components of the vector remains invariant. A 
matrix obeying the first two conditions only is a {\it stochastic} 
matrix; if a discrete probability distribution is thought of as a 
vector $\vec{p}$ then the vector $\vec{q} = B\vec{p}$ is a probability 
distribution too. The third condition ensures that the uniform 
distribution, a vector all of whose components are equal, is transformed 
into itself. Hence a bistochastic matrix causes a kind of contraction 
of the probability simplex with the uniform distribution as a fixed 
point. 

One way of obtaining a bistochastic matrix is to start with a unitary 
matrix $U$ and take the absolute value squared of its matrix elements,

\begin{equation} B_{ij} = |U_{ij}|^2 \ . \label{2} \end{equation} 

\noindent If there exists such a $U$ then $B$ is said to be 
{\it unistochastic}. This raises two mathematical questions:

{\bf I}: Given a bistochastic matrix, is it unistochastic?

{\bf II}: If so, to what extent is $U$ determined by $B$?

\noindent My first task is clearly to convince you that these questions 
are interesting.

Indeed these questions occur in several approaches to quantum foundations. 
An early example is that of Alfred Land\'e \cite{Lande}. More recent examples 
include those of Carlo Rovelli \cite{Rovelli} and Andrei Khrennikov 
\cite{Khrennikov}. Roughly speaking the reason is 
that one first argues that transition probabilities, suitably defined, form 
bistochastic matrices. In attempting to build some group structure into 
these transition probabilities one is then led to require that they form 
unistochastic matrices, and the interference structure that is typical of 
quantum mechanics follows. But here our questions I and II are clearly 
relevant. 

Particle physicists form another set of people interested in unistochastic 
matrices. Here question II is at the center of interest. Thus in the theory 
of weak interactions we encounter the unitary Kobayashi--Maskawa matrix, 
and Cecilia Jarlskog raised the question whether the difficult to measure 
phases in this matrix---that contain CP--violating effects in the theory---can 
be obtained by measuring the moduli of the matrix---that correspond to 
easily measured decay rates. Up to ``rephasing'' (to be 
explained later) it turns out that $B$ determines $U$ uniquely except for 
a discrete ambiguity for $3 \times 3$ matrices \cite{Jarlskog}, while this 
is not so for $4 \times 4$ \cite{Auberson}. As far 
as the KM matrix is concerned $N = 3$ is the interesting case; $N = 4$ was 
studied just in case a fourth generation of quarks should be discovered. 
(The same question occurs in scattering theory, and there no restriction on 
$N$ is imposed \cite{Mennessier}.)

Returning to quantum mechanics proper, there are various corners of quantum 
information/computation theory where questions concerning unistochastic matrices 
arise. My own interest came from an attempt to sharpen Schr\"odinger's mixture 
theorem (on the various ways that a given mixed state can be represented as a 
mixture of pure states) \cite{Bengtsson}. Another example has to do with 
quantum mechanics on graphs \cite{Tanner} \cite{Pakonski}. In this 
connection studies of the spectra and entropies of unistochastic matrices chosen at 
random have been made \cite{Karol}. In these applications question I comes to 
the fore again---in the second application rephrased as a question about what 
Markov processes that have a quantum counterpart in the given context. I will 
give further examples later when I have introduced a little more terminology.  


Assuming that questions I and II are now on the table, let us begin 
by discussing the structure of the set ${\cal B}_N$ of bistochastic $N \times N$ 
matrices. It is a convex polytope called {\it Birkhoff's polytope}, a 
structure well known in the theory of linear programming. Its dimension is 
$(N-1)^2$ and its corners are the $N!$ permutation matrices \cite{Birkhoff}. 
For $N = 2$ the set is just a line segment, while for $N = 3$ we have a 
four dimensional polytope with six corners. We can easily draw its graph, 
that is to say we draw all its corners and all its extremal edges. It turns 
out that all its edges are extremal, which is a rather exceptional property---in 
three dimensions only the simplex has this property. 


For all $N$ Birkhoff's polytope is centered at the {\it van der Waerden matrix}, 
all of whose matrix elements are equal. It is called that because van der Waerden 
made some conjecture about it. The van der Waerden matrix is always unistochastic. 
A corresponding unitary is the Fourier matrix, whose matrix elements are 

\begin{equation} U_{ij} = q^{ij} \ , \hspace{1cm} 0 \leq i,j \leq N-1 \ , 
\end{equation}

\noindent where $q = e^{2{\pi}i/N}$ is a root of unity. In general a unitary 
matrix giving rise to the van der Waerden matrix through eq. (\ref{2}) is known 
as a {\it complex Hadamard matrix}. If it is real, it is known simply as an 
Hadamard matrix. The task of finding all real Hadamard matrices has occupied 
mathematicians since 1867 (when Sylvester first introduced them \cite{Sylvester}); 
the problem 
is of considerable interest to computer scientists since Hadamard matrices 
are useful for constructing error correcting codes, and in other ways. Hadamard 
observed that they can exist only if $N = 2$ or $N = 4k$ and conjectured that 
they do exist in these dimensions \cite{Hadamard}. His conjecture has proved 
a hard nut to crack \cite{Hedayat}. 
In quantum information theory the restriction to real Hadamard matrices is not 
natural. An example of the usefulness of complex Hadamard matrices is provided 
by the fact that they can be used to construct bases of maximally entangled vectors. 
This in turn is an interesting problem because it is known that the set of 
maximally entangled bases is in one-to-one correspondence to the set of dense 
coding schemes, or equivalently teleportation schemes \cite{Werner}. Let us 
see how the construction goes, choosing a Hilbert space of dimension $3 \times 3$ 
as an illustration. Choose a basis $|1\rangle$, $|2\rangle $, $3\rangle $ in each 
factor Hilbert space. Write down the nine vectors 

\begin{eqnarray} |1\rangle |1\rangle + |2\rangle |2\rangle + |3\rangle |3\rangle , 
\ |1\rangle |1\rangle + q|2\rangle |2\rangle + q^2|3\rangle |3\rangle , 
\ |1\rangle |1\rangle + q^2|2\rangle |2\rangle + q|3\rangle |3\rangle , 
\nonumber \\
|1\rangle |2\rangle + |2\rangle |3\rangle + |3\rangle |1\rangle , 
\ |1\rangle |2\rangle + q|2\rangle |3\rangle + q^2|3\rangle |1\rangle , 
\ |1\rangle |2\rangle + q^2|2\rangle |3\rangle + q|3\rangle |1\rangle , 
\nonumber \\
|1\rangle |3\rangle + |2\rangle |1\rangle + |3\rangle |2\rangle , 
\ |1\rangle |3\rangle + q|2\rangle |1\rangle + q^2|3\rangle |2\rangle , 
\ |1\rangle |3\rangle + q^2|2\rangle |1\rangle + q|3\rangle |2\rangle .
\nonumber \end{eqnarray}

\noindent This is a generalization of the Bell basis in dimension $2 \times 2$, and 
all its members are maximally entangled by construction. That they form a basis 
is also evident. What we have done is to first write down a Latin square, and then 
to insert phases from the Fourier matrix to increase the number of orthogonal vectors 
from $3$ to $3 \times 3$. It is clear that the same construction will work whatever 
the value of $N$ we choose, and whatever Latin square and whatever complex Hadamard 
matric we take. Thus the problem of classifying all dense coding schemes is at least as 
difficult as the problem of classifying all complex Hadamard matrices, plus the 
problem of classifying all Latin squares (a problem that we will not go into here). 

The quantum optics community has also payed attention to complex Hadamard 
matrices \cite{Stenholm}. They are sometimes referred to as {\it Zeilinger 
matrices} due to some scheme with symmetric multiports proposed by Zeilinger and 
collaborators \cite{Zeilinger}. 

\section{Some modest results}

\noindent Let us now take up question II in some earnest. It is clear that 
uniqueness cannot hold. Let $D_1$ and $D_2$ be diagonal unitary matrices. Then it 
is clear that $U$ and 

\begin{equation} U' = D_1UD_2 \end{equation}

\noindent will give rise to the same bistochastic matrix $B$ under eq. (\ref{2}). 
The most we can hope for is a one-to-one correspondence between the set of 
unistochastic matrices and the double coset space 

\begin{equation} U(1)\times \dots \times U(1)\setminus U(N)/U(1)\times \dots \times 
U(1) \ , \end{equation}

\noindent with $N$ $U(1)$ factors on the right and $N -1$ factors on the left, say. 
The dimension of this space is the dimension of the set of unitaries minus 
$2N - 1$, that is $N^2 - (2N-1) = (N-1)^2$, which is the dimension of the set 
of bistochastic matrices. There is a slight problem in that the left action 
on the right coset space has fixed points, so our double coset space is not 
smooth. It is easy to locate the fixed points though, so that one 
can treat eq. (\ref{2}) as a map between two smooth manifolds for most 
practical purposes. In practice the phase ambiguity is used to choose the 
first row and the first column to be real and positive. In this way we obtain 
what, in the particle physics community, is known as the set of dephased unitaries, 
and a preliminary conjecture might be that there is a one-to-one correspondence 
between the set of dephased unitaries and the set of bisthochastic matrices. 
When $N = 2$ this conjecture is true. 

For $N > 2$ it is false. For $N = 3$ a dephased unitary can be written as 

\begin{equation} \left[ \begin{array}{lll} r_{00} & r_{01} & \bullet \\ 
r_{10} & r_{11}e^{i{\phi}_{11}} & \bullet \\ r_{20} & r_{21}e^{i{\phi}_{21}} 
& \bullet \end{array} \right] \ . \end{equation}

\noindent The moduli $r_{ij}$ are given (as square roots of matrix elements of 
a bistochastic matrix), and the question is whether one can find phases ${\phi}_{ij}$ 
such that this matrix is unitary. To do so it is enough to check that the first 
two columns are orthogonal; the last column will work out automatically which is 
why we do not write it explicitly. The problem is equivalent to that of choosing 
two angles so that three given lengths form a triangle. This may or may not 
be possible, depending on the lengths. (There are altogether six such ``unitarity 
triangles'' associated to our matrix. Interestingly they all have the same area 
\cite{Jarlskog}.)


Since the answer to question I is sometimes yes, sometimes no, the question 
becomes that of understanding the set of unistochastic matrices as a subset 
of ${\cal B}_3$. To do this we visualize the polytope, or at least we 
organize our impressions of it, with the observation that its 3 + 3 corners 
form the vertices of two equilateral triangles that sit in two orthogonal 
2-planes. Then 
we look at one of these triangles and see how the unistochastic 
subset sits in that triangle. Detailed study confirms the impression we get from 
this picture. The salient facts are that there is a ball of unistochastic 
matrices surrounding the van der Waerden matrix, and then there is a ``spiky'' 
structure which hits the boundary of the polytope in a two dimensional set. 
(Incidentally the unistochastic set has codimension 1 in the boundary for all $N$.) 
Technically the unistochastic set is star shaped and its relative volume is 
(numerically) close to 75 percent. Its boundary consists of 
orthostochastic matrices, that is bistochastic matrices for which 
the matrix $U$ can be taken to belong to the orthogonal group. The map from 
the set of dephased unitaries is generically two to one, so the answer to 
question II is that there is a discrete ambiguity. 

The story for $N = 4$ is much richer. Birkhoff's polytope is now nine 
dimensional and has $24$ corners, which form altogether 6 regular tetrahedra. 
It is no longer true that all edges are extremal. The 2-dimensional faces consist 
of triangles and squares. (Incidentally this is true for any $N$ \cite{Brualdi}.) 
There are 18 squares altogether and their diagonals are precisely the edges of the 
regular tetrahedra. The polytope turns out to be organized around 
nine orthogonal hyperplanes, each containing the corners of four of 
the tetrahedra. Moreover each tetrahedron contains the normal vectors of three 
of the hyperplanes. 


Questions I and II now become calculationally difficult to answer. 
Indeed very much so; an attempt to check whether a given bistochastic 
matrix is unistochastic using either a direct attack, or else some 
parametrization of unitary matrices, typically leads to algebraic 
equations of very high orders. So we are stuck with a difficult problem 
in algebraic geometry. It was pointed out already by Hadamard \cite{Hadamard} 
that continuous ambiguities appear 
in the answer to question II. In fact the most general complex Hadamard matrix 
is (up to permutations of rows and columns)  

\begin{equation} \frac{1}{2} \left[ \begin{array}{cccc} 1 & 1 & 1 & 1 \\ 
1 & e^{i{\phi}} & - 1 & - e^{i{\phi}} \\ 1 & - 1 & 1 & -1 \\ 
1 & -e^{i{\phi}} & - 1 & e^{i{\phi}} \end{array} \right] \ . 
\label{10} \end{equation}

\noindent In a calculational {\it tour de force}, Auberson, Martin and 
Mennessier \cite{Auberson} were able to determine all bistochastic $4\times 4$ 
matrices for which there are such continuous ambiguities in the answer to 
question II. Their occurence clearly complicates matters. Detailed calculations 
shows that at a complex Hadamard matrix the tangent space of the set of dephased 
unitaries maps to one of the nine orthogonal hyperplanes around which the global 
structure of the polytope is organized; what we are looking at is a kind of nine 
dimensional snowflake where the center determines the periphery, and conversely. 
On the face of it there are two explanations for the degeneracy: Either we have 
the rather standard situation known from the ``blow up of a point'' in 
algebraic geometry, where the map from the set of dephased unitaries fails to 
be one-to-one at some isolated points, or, more dramatically, the van der Waerden 
matrix actually lies at the boundary of the unistochastic set. Unfortunately 
I do not know which is the case, but the evidence so far (from numerics and 
computer algebra) rather favours the latter explanation.\footnote{The latter 
explanation has since been proved to be the correct one \cite{IB}.} 


About higher dimensions not much is known, although we do have evidence 
that the set of unistochastic matrices always has the full dimension 
$(N-1)^2$. For prime $N$ there is always a unistochastic ball around the 
van der Waerden matrix. When $N$ is not prime the situation is again 
unclear since the image of the tangent space at the Fourier matrix 
degenerates \cite{Tadej}. Complex 
Hadamard matrices have been looked for for modest values of $N$. There 
are some standard methods to produce them, such as using the character 
table of some finite group. (Choosing the cyclic group this gives 
exactly the Fourier matrix.) Further, typically non-equivalent, examples 
can be found if one first finds what is known as a bi--unimodular sequence, 
and then forms the circulant matrix of the sequence. The first examples 
of such sequences were found by Gauss, and all examples have been 
classified by G\"oran Bj\"orck up to $N = 8$ \cite{Bjorck}. Examples of complex Hadamard 
matrices not coming from any of these two methods have been found for 
$N =6$; continuous ambiguities appear when $N = 4$, $6$ and $8$ while 
the solution for $N = 5$ is unique up to permutations of rows and columns. 
Continuous ambiguities also appear for some prime $N$, although Petrescu 
\cite{Petrescu} has proved that the Fourier matrix is always an isolated 
point when $N$ is prime. For further information on this subject I recommend 
the papers by Haagerup \cite{Haagerup} and by D\u{i}\c{t}a \cite{Dita}.

\section*{Acknowledgments}

\noindent I thank my collaborators on unistochastic matrices, \AA sa Ericsson, 
Marek Ku\'s, Wojciech S{\l}omczy\'nski, Wojciech Tadej and Karol \.Zyczkowski, 
for our many discussions. Unacknowledged results mentioned here are due to them; 
some have been reported in a joint publication \cite{IB}. 
I also thank G\"oran Bj\"orck for kindly explaining his results to me, 
Uffe Haagerup for sending a copy of Petrescu's thesis, and of course Andrei 
Khrennikov for sm\aa l\"andsk g\"astfrihet in V\"axj\"o.




\end{document}